\documentclass{article}

\usepackage{microtype}
\usepackage{graphicx}
\usepackage{subfigure}
\usepackage{booktabs} %
\usepackage[section]{placeins}  %
\usepackage{float}
\usepackage{hyperref}

\usepackage[accepted]{icml2025}

\usepackage{amsmath}
\usepackage{amssymb}
\usepackage{mathtools}
\usepackage{amsthm}

\usepackage[capitalize,noabbrev]{cleveref}

\theoremstyle{plain}

\theoremstyle{definition}

\theoremstyle{remark}

\usepackage[textsize=tiny]{todonotes}

\icmltitlerunning{Improving Genomic Models via Task-Specific Self-Pretraining}

\begin{document}

\twocolumn[
\icmltitle{Improving Genomic Models via Task-Specific Self-Pretraining}

\icmlsetsymbol{equal}{*}

\begin{icmlauthorlist}
\icmlauthor{Sohan Mupparapu}{iiit}
\icmlauthor{Parameswari Krishnamurthy}{iiit}
\icmlauthor{Ratish Puduppully}{itu}
\end{icmlauthorlist}

\icmlaffiliation{iiit}{IIIT Hyderabad, India}
\icmlaffiliation{itu}{IT University of Copenhagen, Denmark}

\icmlcorrespondingauthor{Ratish Puduppully}{rapu@itu.dk}

\icmlkeywords{Machine Learning, ICML}

\vskip 0.3in
]

\printAffiliationsAndNotice{}  %

\begin{abstract}
Pretraining DNA language models (DNALMs) on the full human genome is resource-intensive, yet often considered necessary for strong downstream performance. Inspired by recent findings in NLP and long-context modeling, we explore an alternative: self-pretraining on task-specific, unlabeled data. Using the BEND benchmark, we show that DNALMs trained with self-pretraining match or exceed the performance of models trained from scratch under identical compute. While genome-scale pretraining may still offer higher absolute performance, task-specific self-pretraining provides a practical and compute-efficient strategy for building stronger supervised baselines. 
\end{abstract}

\section{Introduction}

DNA language models (DNALMs) \cite{ji2021dnabert,dallatorre2024nucleotide,nguyen2023hyenadna} have emerged as powerful tools for modeling and interpreting genomic sequences, drawing inspiration from advances in natural language processing (NLP). These models are typically pretrained on large-scale genomic corpora—such as the human genome.

Recent benchmarks like BEND \citep{marin2024bend} highlight that while DNALMs perform well on certain tasks, they struggle with long-range dependencies and exhibit inconsistent performance across biologically meaningful benchmarks. At the same time, findings from long-sequence modeling \citep{amos2024never} suggest that self-pretraining—pretraining on the downstream task data itself—followed by supervised fine-tuning can outperform training from scratch, particularly for long-context tasks.

Motivated by these observations, we investigate whether self-pretraining on task-specific genomic data can yield stronger supervised models under limited compute. Using the BEND benchmark, we find that models trained with task-specific self-supervision match or exceed the performance of scratch-trained counterparts across multiple tasks. While not a substitute for genome-scale pretraining in all settings, our results show that self-pretraining is a compute-efficient strategy for improving downstream performance and building stronger supervised baselines.

This approach is particularly appealing in genomics, where sequences for a given task are abundant but labels are often scarce. By leveraging task-specific unlabeled data, self-pretraining enables models to learn domain-relevant patterns before fine-tuning, improving performance with minimal overhead.

We release the code, pretrained model and finetuned models to support reproducibility\footnote{\url{https://github.com/SohanMupparapu/dna_spt}}.

\section{Related Work}

A growing body of work has applied language modeling to DNA sequences, resulting in several pretrained genomic models. DNABERT \citep{ji2021dnabert} introduced masked language modeling on k-mer tokenized data, showing utility for motif discovery and promoter prediction. Later models like the Nucleotide Transformer \citep{dallatorre2024nucleotide} and HyenaDNA \citep{nguyen2023hyenadna} scaled to larger genomes and longer contexts, improving performance on regulatory genomics tasks. GENA-LM \citep{fishman2023genalm} uses efficient Transformer variants to model full-length sequences. Caduceus \citep{schiff2024caduceus} further improves long-range modeling by incorporating reverse complement symmetry and scalable bidirectional blocks based on Mamba.

Beyond these pretrained models, recent work across NLP and biology has begun to question the necessity of large-scale pretraining for strong downstream performance. \citet{krishna-etal-2023-downstream} demonstrate that language models pretrained solely on downstream data can match or exceed models trained on large corpora such as BookWiki for tasks like classification and QA. Similarly, \citet{amos2024never} show that in long-sequence modeling, self-pretraining on task data substantially narrows the gap between standard Transformers and more complex architectures like S4 \citep{gu2022efficiently} on Long Range Arena tasks \citep{tay2021long}, suggesting that task-specific pretraining can serve as an effective inductive prior.

In the biological domain, \citet{xu2025specialized} report that foundation models (FMs) in genomics often fail to outperform well-tuned supervised baselines, despite significantly higher computational costs. Supporting this, \citet{vishniakov2024foundationless} show that randomly initialized models can match or even outperform pretrained genomic models across tasks like enhancer and histone mark prediction. 
Our work builds on these insights in the context of DNA language modeling by focusing on strengthening supervised baselines through self-pretraining—an approach that is both compute-efficient and independent of large upstream corpora.

\section{Tasks in BEND}
We evaluate on four tasks from the BEND benchmark~\cite{marin2024bend} (see Appendix Table~\ref{tab:appendix-bend} for an overview).

\textbf{Gene Finding.} A multiclass sequence labeling task where each nucleotide is labeled as exon, intron, splice site (donor/acceptor, forward/reverse strand), or noncoding. Derived from GENCODE \cite{frankish_gencode_2021} annotations, this task evaluates both local and long-range understanding of gene structure. Performance is measured with multiclass MCC \cite{gorodkin2004mcc}.

\textbf{Chromatin Accessibility Prediction.}
A multilabel classification task where models predict DNase I hypersensitivity across 125 human cell types from 512 bp sequences. Labels are derived from ENCODE DNase-seq data \citep{encode2012integrated}, following preprocessing from \citet{kelley2016basset}. Each label indicates whether the chromatin is open in a specific cell type. Evaluation is based on average AUROC across labels.

\textbf{Histone Modification Prediction.}
A multilabel task predicting the presence of 18 histone modifications from 512 bp sequences in the K562 cell line. Labels are based on ENCODE ChIP-seq data for 11 histone marks across 18 replicates \citep{encode2012integrated}. Each label indicates whether a particular mark is bound to the sequence. AUROC is averaged across all histone marks.

\textbf{CpG Methylation Prediction.}
A multilabel classification task to predict the methylation status of CpG sites across 7 human cell lines using 512 bp windows centered on CpG sites. Labels are derived from ENCODE bisulfite sequencing data \citep{encode2012integrated}. AUROC is averaged over all cell lines.

\section{Methods}

\subsection{Model and Pretraining Setup}
We use a residual convolutional neural network (CNN) as a shared encoder across tasks. The architecture, inspired by ResNet~\cite{he2016deep} and the BEND benchmark~\cite{marin2024bend}, consists of 30 convolutional layers (kernel size 9) with 512 hidden channels. Dilation doubles with each layer (reset every 6 layers, max 32), and we use GELU activation and LayerNorm.

During self-supervised pretraining, we attach a masked language modeling (MLM) head to the encoder and train on unlabeled DNA sequences. Inputs are one-hot encoded over a vocabulary of size $V=7$ (\texttt{A}, \texttt{C}, \texttt{G}, \texttt{T}, \texttt{N}, \texttt{[MASK]}, \texttt{[PAD]}). Tokens are masked independently with probability $p_{\mathrm{mlm}}=0.15$, using the standard 80/10/10 replacement strategy. The MLM loss is computed as
$\mathcal{L}_{\mathrm{MLM}} = -\sum_{i: m_i=1} \log P_{\theta}(x_i \mid \widetilde{x})$,
where $m_i$ indicates masked positions and $\widetilde{x}$ is the corrupted input.

\subsection{Supervised Fine-Tuning and Training Configuration}
After pretraining, we replace the MLM head with a task-specific predictor consisting of a two-layer CNN (stride 1, ReLU activation) and a linear output layer. The model is fine-tuned end-to-end on downstream tasks. We use cross-entropy loss for single-label classification (e.g., gene finding) and binary cross-entropy loss for multi-label tasks (e.g., chromatin accessibility, histone modifications, CpG methylation).

All models are implemented in PyTorch and trained on a mix of L40S and A100 GPUs, using code adapted from the BEND benchmark. Pretraining is done on 4,780 gene-finding samples over 20 epochs. Fine-tuning is performed for 10 epochs on gene finding and 5 epochs on the other tasks. The latter are trained for fewer epochs to reduce compute, as they have larger training sets and longer per-epoch runtimes.

\subsection{Pretraining Corpus and Evaluation}
We use gene-finding sequences for pretraining and reuse the resulting encoder across all tasks. These sequences are the longest in the BEND benchmark (1,433–14,000 bp), enabling modeling of long-range dependencies. Although labels are not used during pretraining, the sequences cover biologically structured regions such as exons, introns, and splice sites—features closely linked to regulatory activity and downstream task relevance.

\section{Results}

We evaluate whether self-supervised pretraining on task-specific data (\textit{SPT}) improves downstream performance compared to training from scratch (\textit{scratch}). We begin with a compute-matched comparison using the same ResNet architecture, and then benchmark \textit{SPT} against expert methods and genome-pretrained models from the BEND benchmark.

\paragraph{Compute-Matched Comparison.}
Table~\ref{tbl:results} shows that \textit{SPT} consistently achieves equal or better performance than \textit{scratch} under identical compute. Gains are especially notable for gene finding and CpG methylation, with absolute improvements of 12 and 5 points, respectively. For chromatin accessibility and histone modification, both models perform similarly—suggesting that these tasks are already well-solved by supervised training, leaving limited room for additional improvement via pretraining. Notably, even a shallow CNN (Table~\ref{tbl:final}, row “CNN”) performs competitively on histone modification, indicating that this task may be relatively insensitive to architectural depth or pretraining.

\paragraph{Improving Gene Finding via Sequence Labeling.}
In the BEND benchmark, gene finding is framed as a per-base classification task. To explore whether modeling label dependencies improves performance, we augment our \textit{SPT} model with a neural linear-chain Conditional Random Field (CRF) \cite{lample-etal-2016-neural,lafferty2001conditional}, drawing inspiration from the structured HMM-based approach used in \textsc{Augustus} \cite{rabiner1989tutorial,stanke2003gene}. The CRF layer captures transition dynamics across sequence positions, promoting globally coherent predictions (e.g., valid exon–intron boundaries). Using the \texttt{torchcrf} implementation,\footnote{\url{https://github.com/rikeda71/TorchCRF}} we observe a substantial improvement: MCC increases from 0.50 to 0.64 (Table~\ref{tbl:results}).
This gain suggests that gene boundaries and exon-intron transitions benefit from structured prediction, which enforces global label consistency—something per-position classifiers cannot capture. CRFs provide an inductive bias toward valid biological transitions, which may be especially helpful in low-data settings.

\begin{table}[t]
\caption{Performance of models trained from scratch, with self-pretraining (SPT), and SPT with CRF. CRF improves gene finding by modeling label transitions.}
\vskip 0.15in
\centering
\small
\begin{tabular}{lp{1.2cm}p{1.2cm}p{1.2cm}p{1.2cm}}
\toprule
& \textbf{Gene finding} & \textbf{Chromatin accessibility} & \textbf{Histone modification} & \textbf{CpG Methylation} \\
\midrule
\textit{scratch}         & 0.38 & 0.87 & 0.76 & 0.89 \\
\textit{SPT}             & 0.50 & 0.87 & 0.77 & 0.94 \\
\textit{SPT + CRF}       & 0.64 & -- & -- & -- \\
\bottomrule
\end{tabular}
\label{tbl:results}
\end{table}

\begin{figure}[t]
    \centering
    \includegraphics[width=0.95\linewidth]{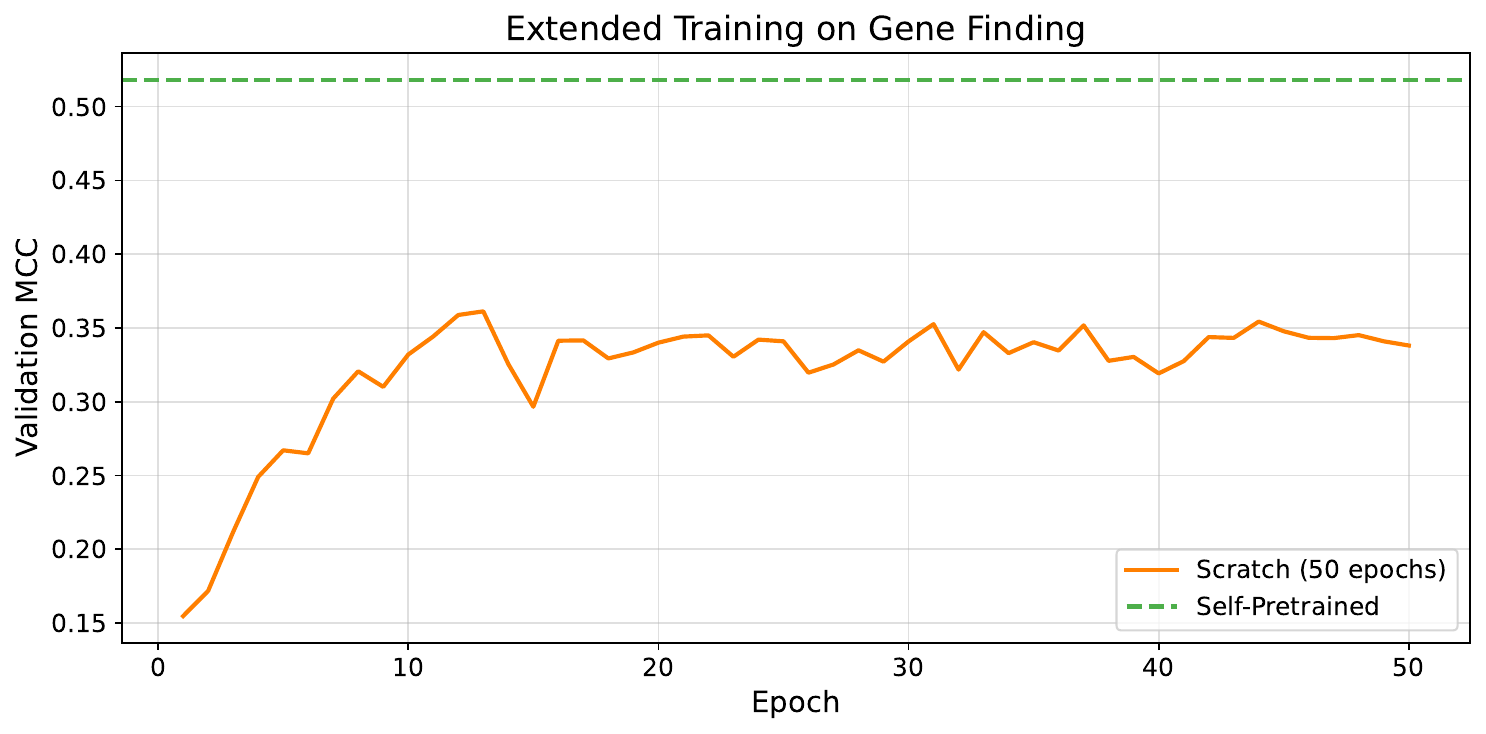}
    \caption{
Validation MCC on gene finding. The scratch model plateaus early and does not match the self-pretrained model.
    }
    \label{fig:gene-finding-extended-training}
\end{figure}

\paragraph{Extended Training Comparison.}
To assess whether longer training improves performance for \textit{scratch} models, we train the \textit{scratch} model on gene finding for 50 epochs—substantially more than the 10 used in our earlier comparison. Our \textit{SPT} model, by contrast, uses 20 epochs of unsupervised pretraining followed by supervised fine-tuning. As shown in Figure~\ref{fig:gene-finding-extended-training}, the  \textit{scratch} model improves initially but plateaus around epoch 13 and fails to match the performance of the \textit{SPT} model. A dashed horizontal line indicates the best validation MCC achieved by the \textit{SPT} model, highlighting the efficiency of self-pretraining over prolonged supervised training.

\begin{figure}[t]
    \centering
    \includegraphics[width=0.95\linewidth]{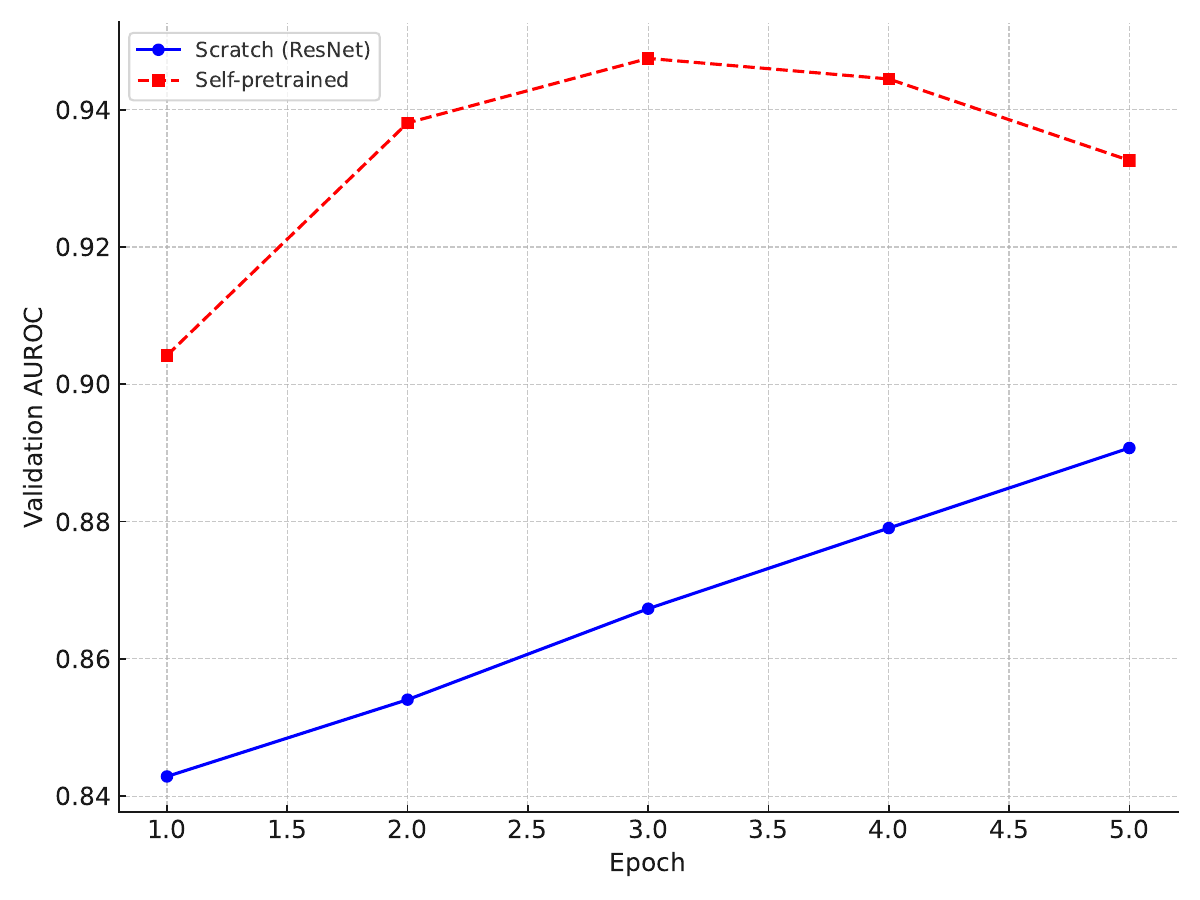}
    \caption{Validation AUROC on the CpG methylation task over training epochs.}
    \label{fig:val-auroc-cpg}
\end{figure}

We now focus on CpG methylation, where \textit{SPT} shows strong gains as well.
\paragraph{Training Dynamics.}
Figure~\ref{fig:val-auroc-cpg} shows validation AUROC over training epochs. The performance gap between \textit{SPT} and \textit{scratch} appears from the first epoch and persists throughout, demonstrating the benefits of task-specific self-pretraining.

\begin{figure}[t]
    \centering
    \includegraphics[width=0.9\linewidth]{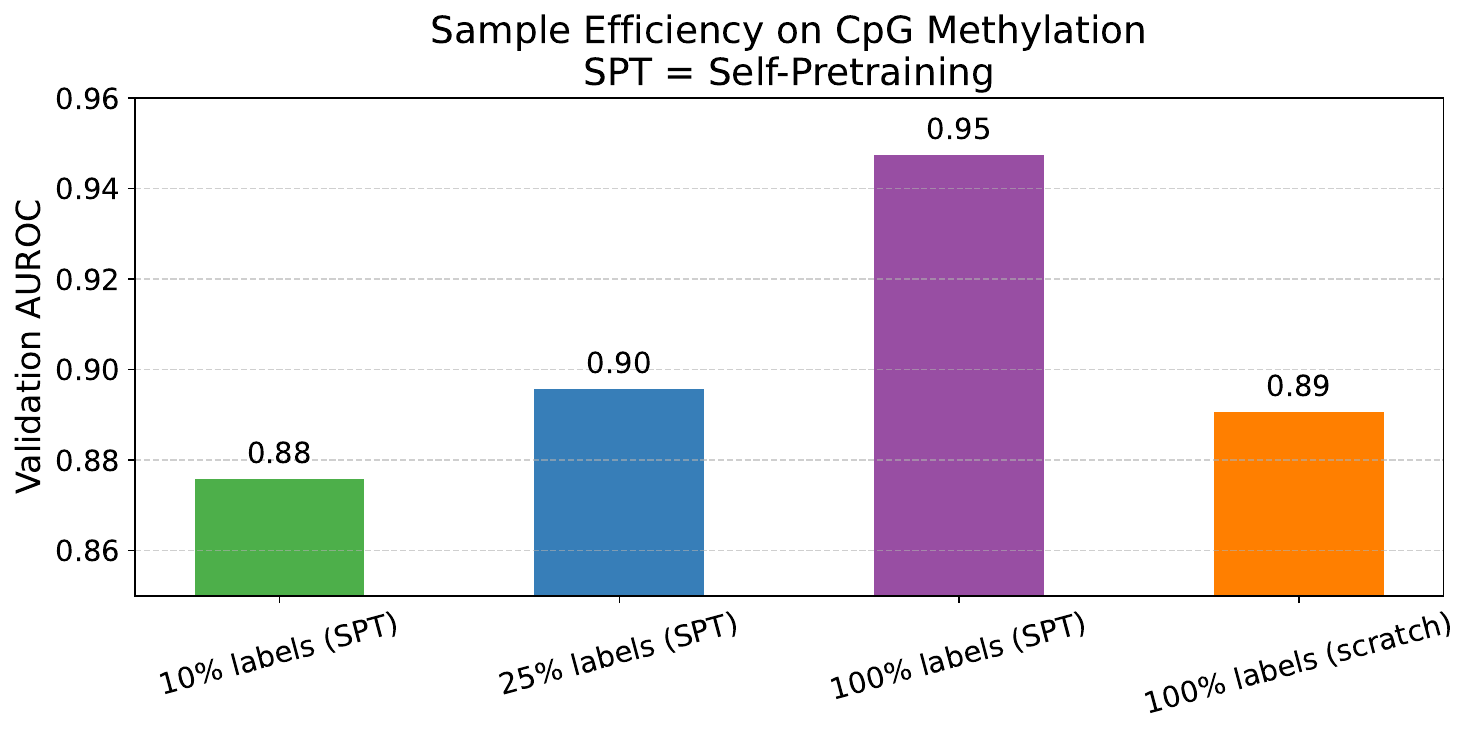}
    \caption{
    Maximum validation AUROC on CpG methylation using different fractions of labeled data. The 25\% self-pretrained (SPT) model outperforms the 100\% scratch model, while the 10\% SPT model performs slightly below.
    }
    \label{fig:sample-efficiency-cpg}
\end{figure}

\begin{table*}[t]
\caption{
Performance on selected BEND benchmark tasks. We compare expert methods, supervised baselines, genome-pretrained DNA LMs, and our self-pretrained (\textit{SPT}) models. \textit{SPT + CRF}$^{\dagger}$ augments the gene finding model with a linear-chain CRF decoder. Best results for each task are shown in bold.
}
\vskip 0.15in
\centering
\small
\begin{tabular}{lcccc}
\toprule
& \textbf{Gene finding} & \textbf{Chromatin accessibility} & \textbf{Histone modification} & \textbf{CpG Methylation} \\
\midrule
\textbf{Expert method} & 0.80 & 0.85 & 0.74 & 0.93 \\
& \textsc{Augustus} & \textsc{Basset} & \textsc{Basset} & \textsc{Basset} \\
\midrule
\textbf{Fully supervised} & & & & \\
ResNet & 0.46 & - & - & - \\
CNN    & 0.00 & 0.75 & 0.76 & 0.84 \\
\midrule
\textbf{Pre-trained} & & & & \\
ResNet-LM         & 0.36 & 0.82 & 0.77 & 0.87 \\
AWD-LSTM          & 0.05 & 0.69 & 0.74 & 0.81 \\
NT-H              & 0.41 & 0.74 & 0.76 & 0.88 \\
NT-MS             & \textbf{0.68} & 0.79 & 0.78 & 0.92 \\
NT-1000G          & 0.49 & 0.77 & 0.77 & 0.89 \\
NT-V2             & 0.64 & 0.80 & 0.76 & 0.91 \\
DNABERT           & 0.20 & 0.85 & \textbf{0.79} & 0.91 \\
DNABERT-2         & 0.43 & 0.81 & 0.78 & 0.90 \\
GENA-LM BERT      & 0.52 & 0.76 & 0.78 & 0.91 \\
GENA-LM BigBird   & 0.39 & 0.82 & 0.76 & 0.91 \\
HyenaDNA large    & 0.35 & 0.84 & 0.76 & 0.91 \\
HyenaDNA tiny     & 0.10 & 0.78 & 0.76 & 0.86 \\
GROVER            & 0.28 & 0.82 & 0.77 & 0.89 \\
\midrule
\textit{SPT}  & 0.50 & \textbf{0.87} & 0.77 & \textbf{0.94} \\
\textit{SPT + CRF}$^{\dagger}$ & 0.64 & -- & -- & --\\
\bottomrule
\end{tabular}
\label{tbl:final}
\end{table*}

\paragraph{Sample Efficiency}

To assess whether self-pretraining improves sample efficiency, we train the \textit{SPT} model on the CpG methylation task using varying fractions of the labeled training data: 10\%, 25\%, and 100\%. The self-pretrained encoder is kept fixed across all fine-tuning settings. As shown in Figure~\ref{fig:sample-efficiency-cpg}, the \textit{SPT} model trained with 25\% of the data already outperforms the \textit{scratch} model trained on 100\%. The 10\% variant performs slightly below \textit{scratch} but still shows strong performance. These results highlight the value of self-pretraining for enabling accurate predictions in low-data regimes.
\paragraph{Comparison to Benchmark Models.}
We also compare our \textit{SPT} model to expert methods and genome-pretrained DNA language models from the BEND benchmark~\cite{marin2024bend}. These include expert tools (e.g., \textsc{Augustus} \cite{stanke2003gene}, \textsc{Basset} \cite{kelley2016basset}), supervised CNNs and ResNets, and large-scale pretrained models such as DNABERT~\cite{ji2021dnabert}, Nucleotide Transformer~\cite{dallatorre2024nucleotide}, GENA-LM~\cite{fishman2023genalm}, and HyenaDNA~\cite{nguyen2023hyenadna}.

Table~\ref{tbl:final} reproduces BEND benchmark results and adds our \textit{SPT} and \textit{SPT + CRF} models. Despite using no genome-scale pretraining, \textit{SPT} outperforms all supervised-from-scratch baselines (even when baselines are trained for up to 100 epochs) and matches or exceeds many genome-pretrained models. On the CpG methylation task, it achieves the highest AUROC overall. On gene finding, further gains are possible by treating the task as structured sequence labeling: adding a linear-chain CRF layer (\textit{SPT + CRF}) boosts performance to 0.64 MCC.%

These results demonstrate that task-specific self-pretraining can serve both as a strong supervised baseline and a practical alternative to genome-scale pretraining—especially under compute constraints. %

\section{Conclusion}
We demonstrate that self-supervised pretraining on task-specific genomic data—using only unlabeled sequences from the gene-finding task—can improve downstream performance compared to models trained from scratch. Across tasks in the BEND benchmark, self-pretraining consistently yields stronger results under matched compute budgets. While it may not always outperform genome-wide pretraining in absolute terms, our approach provides a practical and compute-efficient alternative that establishes a stronger supervised baseline for genomic modeling.

\section*{Impact Statement}

This paper presents work whose goal is to improve the efficiency and accessibility of machine learning for genomics by exploring task-specific self-pretraining. While our methods may have downstream applications in health and biological research, we do not foresee any immediate ethical or societal risks that require specific attention. There are many potential broader impacts of machine learning in genomics, none of which we feel must be specifically highlighted here.

\bibliography{paper}
\bibliographystyle{icml2025}
\newpage
\appendix
\onecolumn
\section*{Appendix}
\begin{table}[ht]
\caption{Overview of four BEND benchmark tasks. Adapted from \citet{marin2024bend}.}
\label{tab:appendix-bend}
\centering
\small
\begin{tabular}{lp{2cm}ccp{2cm}cp{2cm}}
\toprule
\textbf{Task} & \textbf{Type} & \textbf{\# Samples} & \textbf{Length range} & \textbf{Evaluation} & \textbf{Metric} & \textbf{Source} \\
\midrule
Gene finding & Nucleotide-wise Multiclass (9) & 5,976 & 1,433--14,000 bp & 4780/597/597 & MCC & GENCODE \cite{frankish_gencode_2021} \\
Chromatin accessibility & Sequence-wise Multilabel (125) & 2,005,617 & 512 bp & 1,354,042/ 279,422/ 372,153 & AUROC & ENCODE \cite{encode2012integrated} \\
Histone modification & Sequence-wise Multilabel (18) & 612,081 & 512 bp & 70,801/ 120,567/ 743,095 & AUROC & ENCODE \cite{encode2012integrated} \\
CpG methylation & Sequence-wise Multilabel (7) & 959,039 & 512 bp & 109,717/ 106,227 & AUROC & ENCODE \cite{encode2012integrated} \\
\bottomrule
\end{tabular}

\end{table}

\end{document}